\documentclass[jkps,preprint, fleqn,showpacs,showkeys]{revtex4}

\usepackage[pdftex]{graphicx}
\usepackage{amssymb}
\usepackage{bm}
\begin{document}
\setcounter{page}{1}
\title[]{Interacting Superparamagnetism in La$_{0.7}$Sr$_{0.3}$MnO$_{3}$ Nanoparticles}
\author{Ha M. \surname{Nguyen}}
\email{nmha@ess.nthu.edu.tw}
\thanks{Current address: Department of Engineering and System Science, National Tsing Hua Univesity, Hsinchu, Taiwan 13003, R.O.C}
\author{D. H. Manh}
\author{L. V. Hong}
\author{N. X. Phuc}
\affiliation{Institute of Materials Science, VAST, 18 Hoang-Quoc-Viet, Hanoi, Vietnam}
\author{Y. D. \surname{Yao}}
\affiliation{Institute of Physics, Academia Sinica, Nankang, Taipei, Taiwan 11529, R.O.C}
\date{\today}

\begin{abstract}
The magnetic order of La$_{0.7}$Sr$_{0.3}$MnO$_{3}$ nanoparticles (NPs) fabricated in a SPEX D8000 mill was systematically studied. The La$_{0.7}$Sr$_{0.3}$MnO$_{3}$ nanocrystals grow from  the milled constituent oxides during the milling processes. The magnetization data obtained by using a SQUID magnetometer show the NPs as a superparamagnet in terms of anhysteretic curves near room temperature. Unoverlaping of the scaled $M(H_{\rm{ext}},T)/M_{s}$ vs. $H_{\rm{ext}}/T$ plots and the dc susceptibility obeying the Curie-Weiss behavior rather than the Curie law at high temperatures provide evidence that the NPs are interacting superparamagnetic ensembles. A mean-field correction to the Langevin function $\mathcal{L}\left([H_{\rm{ext}}+\alpha M]/k_{\rm{B}}T\right)$  worked well for the magnetic ordering of the NPs.  By means of the Langevin fitting, the diameter of the NP was estimated to be lower than 15 nm, depending on the milling time. The saturation magnetization of NPs varied from 48.5 em/g to 19 emu/g, with the higher value corresponding to a larger particle size. A core-shell structure of the NP was adopted, with the NP having the core-shell magnetically-effective mass density. This is applicable to the variation of the saturation magnetization with particle size. 

\end{abstract}

\pacs{61.46.Df, 67.80.Jd,75.20.-g, 74.25.Ha,}

\keywords{ Superparamagnetism, Maganite nanoparticles, Ball milling}

\maketitle

\section{INTRODUCTION}
The magnetic properties of nanostructured systems are an interesting subject of research for both basic and applications reasons \cite{R1}. In 1946, Kittel \cite{R2} clearly established that a single domain structure would be more energetically stable for particles below a critical size (e.g.,  10$^{-8}$ m for iron), and N\'eel \cite{R3} pointed out that for such small particles, thermal fluctuation prevents the existence of stable magnetization, resulting in a superparamagnetic (SPM) state. The so-called blocking temperature, $T_{\rm{B}}$, separates the SPM state at high temperature from the blocked state at low temperature \cite{R}. For an assembly of identical NPs with volume $V$, the initial susceptibility in the SPM state is given by $\chi_{\rm{sp}}=M_{\rm{s}}^{2}V/3k_{\rm{B}}T$ and has a Curie-law behavior. The magnetization of the assembly is then described by the classical Langevin function $M = M_{\rm{s}}(\coth\xi -1/\xi)$, where  $\xi= M_{\rm{s}}VH_{\rm{ext}}/k_{\rm{B}}T$ and $M_{\rm{s}}$ and $H_{\rm{ext}}$ are the saturation magnetization and the applied field, respectively. Although several attempts have been made to use widely the aforementioned universalities for characterizing NP systems with excellent agreement \cite{R1}, it is worth noting that extreme care should be taken in employing them in certain systems.  Actually, situations have existed in the literature where parameters derived by employing theoretical models for the experimental data deviated greatly from those extracted from real systems \cite{R1}. That deviation was caused by several factors: namely, the broad particle size distribution, the different structural and/or magnetic phases, the local anisotropies, the interparticle magnetic interactions, and the interplay of spins at the surfaces and the inner cores of particles \cite{R1}. As a whole, it is extremely difficult to distinguish which factors play a dominant role in the magnetic behavior for these complicated situations.

 In the present report, we show a situation of superparamagnetic NP systems in which the experimental data from magnetization measurements indicate a strong interaction among particles, leading to lack of collapsing of the conventional scaled anhysteretic curves into a universal Langevin function. The NP systems addressed in this report are perovskite La$_{0.7}$Sr$_{0.3}$MnO$_{3}$ (LSMO) NPs fabricated from initial oxide constituents by using a reactive milling method. The material is well known to possess the highest Curie temperature of about 375 K and the highest saturation magnetization of 103 emu/g \cite{R4} among the La$_{1-x}$Sr$_{x}$MnO$_{3}$ colossal magnetoresistance perovskite family in bulk form. It has also been shown that because of the nature of the ball milling process, a nonmagnetic surface layer or disordered spin orientation exists at the surfaces of the NPs, which leads to quite different contributions to the particle magnetization \cite{R5}. For a suitable magnetic description of the milled NPs, we used shell-core-type magnetic NPs proposed by Gangopadhyay {\it et al.} \cite{R6}. Furthermore, the interparticle interaction was well described by using the mean-field approximation.


\section{EXPERIMENTAL}
As reported previously in Ref. 8, the powders of LSMO nanocrystallites were produced by milling the mixture of the constituent oxides for various times from 4 to 24 h. The milling processes were performed by using a commercial mill (SPEX D8000 Mixer/Mill) with a combination of two 1/2-inch and four 1/4-inch steel mill balls. The structures of the nanocrystallites were characterized on an X-ray diffractometer (SIEMENS D5000) and showed that the nanocrystal of the LSMO perovskite phase had already formed after 4 h of milling. To study the magnetic properties of the NPs, we chose the milling time to be longer than 8 h to ensure that the powders entirely consist of the LSMO-phase NPs (see the milling-time-dependent X-ray patterns for the LSMO powders in Fig. 1(b) of Ref. 8). The mean grain size and the grain size distributions of the systems were determined by means of various techniques. A commercial WIN-CRYSIZE program packet based on the Warren–Averbach formalism was used to analyze the XRD data. The transmission electron microscope (TEM) measurements were done on a Philip CM20-FEG machine at 200 kV. The mean grain size deduced from the XRD data ($\langle D_{\rm{X-ray}}\rangle$) and from the TEM micrographs ($\langle D_{\rm{TEM}}\rangle$) shown in table 1 indicated that LSMO particles had nanometer diameters that decreased with increasing the milling time. The disagreement between the grain size from the two methods, as already explained in Ref. 8, is due to the structural domain formation in individual nanoparticles, which points out the limitation of the Warren-Averbach method in estimating the size of such systems. It is interesting to show previously in Ref. 8 that SEM images taken on a JSM-5410LV system showed a cluster formation of nanoparticles. The nanoparticles inside each cluster can be observed using high-solution TEM. This observation is important for demonstrating the strong dipolar interaction between particles. For studying the magnetic properties of the perovskite nanoparticles, we used a MPMS5 SQUID magnetometer for the dc magnetic measurements in the temperature range of 5-400 K for fields up to 50 kOe.

 \section{MAGNETIC PROPERTIES}
\begin{figure}[t!]
 \includegraphics[width=5.0cm]{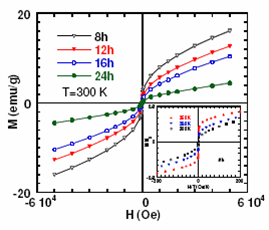}
 \caption{Anhysteretic curves for NPs with different
milling times measured at $T$=300 K. Inset: $M/M_{\rm{p}}^{\rm{s}}$
vs. $H_{\rm{ext}}/T$ curves measured at different temperatures the sample with an 8-h milling.} \label{fig:1}
\end{figure}

\begin{figure}[t!]
 \includegraphics[width=5.0cm]{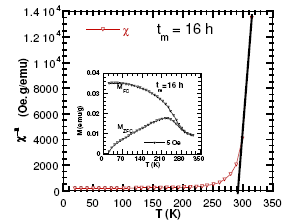}
 \caption{Inverse dc susceptibility vs.
temperature measured at 5 Oe for the sample with
a 16-h milling. Inset: the $M_{\rm{zfc}}$ and the $M_{\rm{fc}}$ curves measured at 5 Oe for the sample.} \label{fig:2}
\end{figure}

Figure \ref{fig:1} presents $M(H)$ curves measured at 300 K for all samples with different milling times.  As seen in the figure, the anhysteresic behavior is observed for all samples. The behavior is also observed for $M(H)$ curves measured at temperatures in the range of 250 K to 370 K. This indicates that the NP ensembles are superparamagnetic in this range. It is worth noting that the scaled plots of $M/M_{\rm{s}}$ vs. $H_{\rm{ext}}/T$ for each sample do not overlap into a universal magnetization curve as desired for noninteracting SPM systems (see, e.g., the inset of Fig. \ref{fig:1} for the sample with an 8-h milling). The reason for this deviation is the dipolar interaction between NPs, which leads to a collective contribution to the magnetization. Such a system is termed an \textit{interacting superparamagnet}.  Moreover, this behavior is evident in the thermomagnetization measurements. The dc susceptibility deduced from $M_{\rm{zfc}}(T)$ in an external field of 5 Oe (the inset of Fig. \ref{fig:2}) obeys a Curie-Weiss law, $\chi= C_{\rm{sp}}/(T-T_{\rm{o}})$, rather than a Curie law,  $\chi= C_{\rm{sp}}/T$, in the SPM state for all samples, where $C_{\rm{sp}}$ is a Curie-like constant and $T_{\rm{o}}$ is the intersection of the linearly extrapolated line of $1/\chi (T)$ in the SPM region with the $T$-axis, as shown in Fig. \ref{fig:2} for the 12-h sample. We see that $T_{o}$ takes positive and large value for all samples.

\begin{figure}[t!]
 \includegraphics[width=5.0cm]{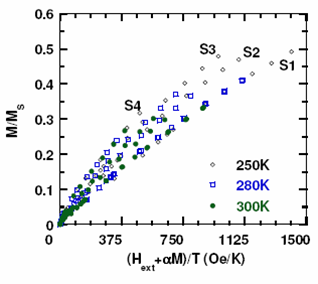}
 \caption{Scaled magnetization curves ($M/M_{\rm{p}}^{\rm{p}}$
versus $(H_{\rm{ext}}+\alpha M)/T)$. (S$_{1}$: 8 h, S$_{2}$: 12 h, S$_{3}$: 16 h, S$_{4}$: 24 h).} \label{fig:3}
 \end{figure}

By means of the law of approach to saturation, the saturation magnetization ($M_{\rm{p}}^{\rm{s}}$) was determined from the plots of $M$ vs. $1/H^{2}$  by extrapolating the value of the magnetization measured at 4.5 K to infinite fields. As shown in table 2, $M_{\rm{p}}^{\rm{s}}$ was found to decrease with increasing milling time. Magnetization values between 48.5 emu/g and 18.9 emu/g were obtained as the milling time was varied from 8 h to 24 h, respectively. The magnetization curves of such interacting SPMs can be well described by using the mean-field approximation with a field correction to the argument of noninteracting SPM Lagevin function added. The magnetization $M$, therefore, can be expressed by an {\it argument-corrected Langevin function}
\begin{equation}
M(H_{\rm{ext}},T) = M_{\rm{p}}^{\rm{s}}\times \mathcal{L}\left(\frac{M_{\rm{p}}^{\rm{s}}\bar{m}[H_{\rm{ext}}+\alpha M(H_{\rm{ext}},T)]}{k_{\rm{B}}T}\right),
\label{eq1}
\end{equation}
where $\mathcal{L}(x)$ is the Langevin function of $x$, $\bar{m}$ is the average particle mass (in g/cm$^{3}$), and $\alpha$ is a parameter proportional to the mean field coefficient $\omega$ ($\omega$ being equal to $\alpha$ devided by the particle density $\rho_{\rm{p}}$). From Eq. (\ref{eq1}), $M(H_{\rm{ext}},T)/M_{\rm{p}}^{\rm{s}}$ is a universal function of $H_{\rm{ext}}+\alpha M(H_{\rm{ext}},T)/k_{\rm{B}}T$. The two parameters $M_{\rm{p}}^{\rm{s}}$ and $\alpha$ are, therefore, required for the scaling process to obtain universal plots. The former was determined as above. The latter was determine by using the experimental data for the initial susceptbility $\chi$ at low field. From Eq. (\ref{eq1}), $\chi$ is expressed as the initial slope of the universal curve,

\begin{equation}  
         \left\{\begin{array}{ll}
                  \chi = \frac{{M_{\rm{p}}^{\rm{s}}}^{2}\bar{m}}{3k_{\rm{B}}(T-T_{\rm{o}})}=
                      \frac{C_{\rm{sp}}}{T-\alpha C_{\rm{sp}}}\\                               

                  \frac{1}{\chi}=\frac{1}{C_{\rm{sp}}}(T-T_{\rm{o}})= \frac{1}{C_{\rm{sp}}}T-\alpha 
               \end{array}\right.,
         \label{eq2}
  \end{equation}
where $C_{\rm{sp}}={M_{\rm{p}}^{\rm{s}}}^{2}\bar{m}/3k_{\rm{B}}$ is a Curie-like constant. We expect a linear relationship between the inverse of the susceptibility and the temperature in the SPM region so that $C_{\rm{sp}}$, $T_{\rm{o}}$, and $\alpha$ can be determined by extrapolating the $1/\chi$ vs $T$ curves as shown Fig. \ref{fig:2} and by using Eq. (\ref{eq2}). $\alpha$ and $M_{\rm{p}}^{\rm{s}}$ are then used to plot the scaled curves of $M/M_{\rm{p}}^{\rm{s}}$ vs. $(H_{\rm{ext}}+\alpha M)/T$ at different temperatures for all samples. As expected, the scaled curves overlap into a universal curve characteristic for each sample. The universal scaling magnetization curves for the experimental data of all samples are shown in Fig. \ref{fig:3}. It is straightforward to think that the argument-corrected Langevin function given in Eq. (\ref{eq1}) should fit the universal curve to determine the average particle diameter $\langle D \rangle$ for each sample.  Actually, $\langle D \rangle$ deduced from such a direct fit is much smaller (about 2-4 nm) than that estimated by using the TEM and the XRD methods as reported in Ref. 8. To understand the disagreement and to estimate $\langle D \rangle$ from the magnetic measurements, a quantitave analysis of the particle structure is addressed. 

As forementioned, the decrease in the saturation magnetization with increasing milling time is due to the decreasing particle size during ball-milling. The reason is assumed to be likely due to the ratio of the magnetization contribution of the surface to that of the volume being higher in smaller particles. As a result, the much higher contribution from the defect and spin-disordered surface layer is released. The values much smaller than 103 emu/g (the bulk saturation magnetization for La$_{0.7}$Sr$_{0.3}$MnO$_{3}$ single crystals) is the evident for the existence of nonmagnetic shells (dead layer) surrounding NP cores \cite{R5}. The shells are poorly magnetic due to noncollinear spin structures or spin-glass-like disorders.  A model of the core-shell structure for small particles is known to be suitable for Fe nanoparticles \cite{R6} and ball-milled nanoparticles of La$_{0.8}$Sr$_{0.2}$MnO$_{3}$ \cite{R5}. In this paper, the model is possible as a phenomenological hypothesis in the magnetic study of La$_{0.7}$Sr$_{0.3}$MnO$_{3}$ nanoparticles. The La$_{0.7}$Sr$_{0.3}$MnO$_{3}$ particle is assumed to be spherical and to be composed of an ideal single-crystalline core with a diameter of $D_{\rm{C}}=2r$. The saturation magnetization of the core is $M_{\rm{C}}^{\rm{s}}$= 103 emu/g, and the density is $\rho_{\rm{C}}$= 6.11 g/cm$^{3}$ (such values corresponds the values of bulk single crystals). The outer layer has a thickness of $dr$ ($dr \ll r$), and the corresponding $M_{\rm{S}}^{\rm{s}}$ and  $\rho_{\rm{S}}$ values are roughly assigned 15 emu/g  and 4 g/cm$^{3}$, respectively. The following expressions [Eq. (\ref{eq4}) and Eq. (\ref{eq5})] can be found from the linear weighted average of the magnetization [Eq. (\ref{eq3})]:
\begin{equation}
M_{\rm{p}}^{\rm{s}}m_{\rm{p}}=M_{\rm{C}}^{\rm{s}}m_{\rm{C}}+M_{\rm{S}}^{\rm{s}}m_{\rm{S}},
\label{eq3}
\end{equation}

\begin{equation}
M_{\rm{p}}^{\rm{s}}=M_{\rm{C}}^{\rm{s}}-3(M_{\rm{C}}^{\rm{s}}-M_{\rm{S}}^{\rm{s}})\frac{\rho_{\rm{S}}}{\rho_{\rm{C}}}\frac{dr}{r},
\label{eq4}
\end{equation}

\begin{equation}
\rho_{\rm{p}}=\rho_{\rm{C}}\times \frac{(M_{\rm{C}}^{\rm{s}})^2-M_{\rm{S}}^{\rm{s}}M_{\rm{p}}^{\rm{s}}}{M_{\rm{p}}^{\rm{s}}(M_{\rm{C}}^{\rm{s}}-M_{\rm{S}}^{\rm{s}})}\times
\left[1+\frac{\rho_{\rm{C}}}{3\rho_{\rm{S}}}+\frac{(M_{\rm{C}}^{\rm{s}}-M_{\rm{p}}^{\rm{s}})}{(M_{\rm{C}}^{\rm{s}}-M_{\rm{S}}^{\rm{s}})}\right]^{-3}.
\label{eq5}
\end{equation}

It is clear from Eq. (\ref{eq4}) that the variation of particle saturation magnetization with respect to $dr/r$ and the nature of core-shell-type particle result in a magnetic and structural dependence of the particle density in Eq. (\ref{eq5}). Therefore, the average mass $\bar{m}$ is related to average particle diameter as follows:
\begin{equation}
\bar{m}=\frac{\pi}{6}\langle D \rangle^{3}\rho_{\rm{p}}.
\label{eq6} 
\end{equation}

\begin{figure}[t!]
 \includegraphics[width=5.0cm]{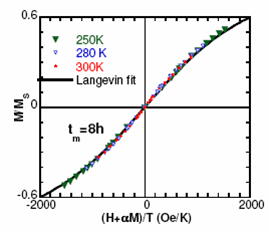}
 \caption{Scaled magnetization curves for the sample with an 8-h milling. Solid line: the magnetically-weighted argument-corrected Langevin function fitting to the scaled data.} \label{fig:4}
 \end{figure}

\begin{figure}[t!]
 \includegraphics[width=5.0cm]{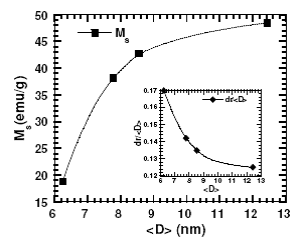}
 \caption{Dependence of saturation magnetization
on average particle size. The inset is the plot of $dr/\langle D \rangle$ vs. $\langle D \rangle$.} \label{fig:5}
 \end{figure}

 To deduce the average particle diameter of each sample, $\langle D \rangle$, we inserted $\bar{m}$ in Eq. (\ref{eq6}) and $\rho_{\rm p}$ in Eq. (\ref{eq5}) into  the argument-corrected Lagevin function in Eq. (\ref{eq1}), then called a {\it magnetically-weighted argument-corrected Langevin function}, and then fit the function with the data of corresponding universal magnetization curve shown in Fig. \ref{fig:3} for each sample.  The estimated values of $\langle D \rangle$ for all samples can be shown in table 2. As one can see, $\langle D \rangle$ varies from 12.4 nm to 6.3 nm with increasing milling time from 8 h to 24 h, respectively. Figure \ref{fig:4} shows an example of a good fit between Eq. (\ref{eq1}) and the experimental data for the sample with an 8-h milling. A plot of the particle saturation magnetization versus the average particle diameter is shown in Fig. \ref{fig:5}. It is clear that $M_{p}^{s}$ increases with particle size and approaches the bulk value for bigger particle sizes. This behavior was also reported by Roy {\it et al.} for ball-milled La$_{0.8}$Sr$_{0.2}$MnO$_{3}$- nanoparticles \cite{R5}. The decrease in the saturation magnetization with increasing milling time (and, hence, decreasing particle size) is likely due to the increasing ratio of the surface thickness to thecore size for decreasing particle size (see the inset of Fig. \ref{fig:5}) as a result of the high-energy milling process.  This was evidenced in our TEM study reported previously\cite{R7}. Finally, once again, the mean-field coefficient $\omega$ obtained by dividing the extrapolated values of $\alpha$ [Eq. (\ref{eq2})] by the corresponding particle density $\rho_{\rm{p}}$ [Eq. (\ref{eq5})] is shown in table 2. $\omega$ is found to be in the range of 175-88, indicating a very strong interaction between particles. 


\section{CONCLUSIONS}
In summary, the magnetic characteristics have been determined for La$_{0.7}$Sr$_{0.3}$MnO$_{3}$ nanoparticles fabricated with different milling times. The analysis of the magnetic measurements using the mean-field correction showed strong interparticle interaction with high values of the mean-field coefficient, which was evidenced by TEM/SEM images and was due to particle aggregation into clusters. The universal scaled magnetization curves in the interacting superparamagnetism state could be described well by an {argument-corrected Langevin function}. Moreover, as to the nature of milling product, the magnetic nanoparticles showed defect nomagnetic surface layers. For a rough magnetic core-shell structure of the particle, the  magnetically-weighted argument-corrected Langevin function fitting yielded a mean particle diameter in the range of 12 nm to 6 nm when the saturation magnetization decreased from 48.5 emu/g to 19 emu/g, respectively. 

\begin{acknowledgments}
This work was supported by grants 4.1.07.07 and 4.1.02.04 of the Priority Nanotechnology Themes in the Fundamental Research Program of Vietnam. 
\end{acknowledgments}


\begin{table}
 \caption{\label{tab1} Particle mean diameter estimated by means of the
XRD and the TEM measurements in Ref. 8.}
 \begin{ruledtabular}
 \begin{tabular}{ccc}
  Milling (hrs) & $\langle D_{\rm{X-ray}}\rangle$ (nm) & $\langle D_{\rm{TEM}}\rangle $ (nm) \\ \hline 
  8  & 11.0 & 17.5 \\
  12 & 8.3  & 12.0 \\ 
  16 & 7.8  & 10.8 \\
  24 & 8.9  & 8.7  \\   
 \end{tabular}
 \end{ruledtabular}
 \end{table}

\begin{table}
 \caption{\label{tab2} Parameters deduced by fitting the data of magnetic
measurements and the magnetically-weighted argument-corrected Langevin function.}
 \begin{ruledtabular}
 \begin{tabular}{cccccc}
  Milling (hrs) & $M_{\rm{p}}^{\rm{s}}$ (emu/g) & $\omega$ & $\rho_{\rm{p}}$ (g/cm$^{3}$) & $\langle D \rangle$ (nm)  \\ \hline 
  8  & 48.49 & 175 & 5.96  & 12.4 \\
  12 & 42.69 & 142 & 6.32  & 8.5  \\
  16 & 38.22 & 140 & 6.69  & 7.9  \\
  24 & 18.89 & 88  & 10.92 & 6.2  \\
 \end{tabular}
 \end{ruledtabular}
 \end{table}



 \end{document}